\documentclass[12pt]{article}

\usepackage[margin=1in]{geometry}
\usepackage{amsmath, amssymb}
\usepackage{graphicx}
\usepackage{booktabs}
\usepackage{array}
\usepackage{multirow}
\usepackage{hyperref}
\usepackage{algorithm}
\usepackage{algpseudocode}
\usepackage{fancyhdr}
\usepackage{titlesec}
\usepackage{caption}
\usepackage{cite}
\usepackage{setspace}
\usepackage{authblk}
\usepackage{rotating}

\title{CBCT-Based Synthetic CT Generation Using Conditional Flow Matching Model}

\author[1]{Junbo Peng}
\author[1]{Huiqiao Xie}
\author[2]{Tonghe Wang}
\author[3]{Xiangyang Tang}
\author[1,*]{Xiaofeng Yang}

\affil[1]{Department of Radiation Oncology and Winship Cancer Institute, Emory University, Atlanta, GA 30322}
\affil[2]{Department of Medical Physics, Memorial Sloan Kettering Cancer Center, New York, NY, 10065}
\affil[3]{Department of Radiology and Imaging Sciences, Emory University, Atlanta, GA 30322}
\affil[*]{Email: xiaofeng.yang@emory.edu}

\date{}

\begin{document}

\maketitle
\thispagestyle{fancy}

% ---- Abstract ----
\begin{abstract}
\noindent\textbf{Background:} Daily or weekly cone-beam computed tomography (CBCT) is employed
in image-guided radiotherapy (IGRT) for precise patient alignment, which is a promising candidate
for adaptive radiotherapy (ART) replanning. However, its clinical utility in quantitative tasks,
including organ segmentation and dose calculation, is hindered by severe artifacts and inaccurate
Hounsfeld unit (HU). To facilitate the implementation of online ART in clinical settings, it is
essential to enhance CBCT image quality to a level comparable with that of conventional CT scans.

\medskip
\noindent\textbf{Purpose:} This purpose of this study is to develop a conditional flow matching
model for synthetic CT generation from CBCT with enhanced image quality.

\medskip
\noindent\textbf{Methods:} The study proposed a conditional flow matching model that gradually
transforms a sample from normal distribution to the corresponding CT sample conditioned on the
input CBCT image. The proposed model was trained using CBCT and deformed planning CT (dpCT)
image pairs in a supervised learning scheme. The feasibility of the conditional flow matching model
was verified using studies of brain, head-and-neck (HN), and lung patients. The quantitative
performance was evaluated using three metrics, including mean absolute error (MAE), peak
signal-to-noise ratio (PSNR), and normalized cross-correlation (NCC). The proposed flow matching
model was also compared to other flow matching and diffusion-based generative models for sCT
generation.

\medskip
\noindent\textbf{Results:} The proposed flow matching model effectively reduced multiple types
of artifacts on CBCT images in all the studies. In the study of brain patient, the MAE, PSNR,
and NCC of the sCT were improved to 26.02 HU, 32.35 dB, and 0.99, respectively, from 40.63 HU,
27.87 dB, and 0.98 on the CBCT images. In the study of HN patient, the metrics were improved to
33.17 HU, 28.68 dB, 0.98 from 38.99 HU, 27.00 dB, 0.98. In the lung patient study, the metrics
were 25.09 HU, 32.81 dB, 0.99 and 32.90 HU, 30.48 dB, 0.98 for sCT and CBCT, respectively.

\medskip
\noindent\textbf{Conclusions:} The proposed conditional flow matching model effectively synthesizes
high-quality CT-like images from CBCT, achieving accurate HU representation and artifact reduction.
This enables more reliable organ segmentation and dose calculation in CBCT-guided online ART
workflows.

\end{abstract}

\newpage

% ---- Table of Contents ----
\tableofcontents

\newpage

% ================================================================
\section{Introduction}
% ================================================================

Cone-beam computed tomography (CBCT) is extensively utilized on a daily or weekly basis in
current image-guided radiotherapy (IGRT) workflows for patient alignment and treatment
monitoring, helping reduce setup errors and enhance treatment precision~\cite{ref1}. However,
compared to conventional multi-detector CT (MDCT), the CBCT images are susceptible to a range
of artifacts---such as beam-hardening induced cupping, shading, streaking and scatter
artifacts---that result in substantial inaccurate Hounsfeld unit (HU) values~\cite{ref2,ref3,ref4}.
These limitations hinder the direct use of CBCT in quantitative applications like organ
segmentation and dose calculation, thus posing a barrier to fully realizing CBCT-based adaptive
radiotherapy (ART). In order to circumvent these issues, current implementations of ART often rely
on deforming the planning MDCT (henceforth pCT) to match CBCT anatomy for dose
calculation~\cite{ref5}. However, the accuracy of deformable image registration is compromised by
CBCT artifacts and anatomical changes during the treatment, making the process heavily dependent
on operator expertise and intuitional protocols~\cite{ref6}.

Recent research aimed at enabling direct CBCT-based ART falls into two major categories: artifact
correction to enhance CBCT image quality~\cite{ref7,ref8,ref9,ref10,ref11,ref12,ref13,ref14,ref15,ref16},
and synthetic MDCT (henceforth sCT) generation from CBCT that approximates the quality of
pCT~\cite{ref6,ref17,ref18,ref19,ref20,ref21}.

In general, CBCT artifact correction methods consist of hardware-, model-, and deep learning-based
approaches. The hardware-based solutions, such as anti-scatter grids~\cite{ref7,ref22},
lattice-shaped beam stoppers~\cite{ref9}, and primary-modulation beam, help suppress scatter
artifacts at a cost of compromised quantum efficiency and degraded signal-to-noise ratio~\cite{ref19}.
The model-based techniques aim to replicate the CBCT imaging physics process to correct artifacts.
For example, Monte-Carlo (MC) simulations~\cite{ref10} and analytical methods~\cite{ref14} have
been developed to estimate scattering signals. While the MC methods provide high accuracy, they
are computationally expensive. The analytical scatter kernel models are more efficient but less
reliable in heterogeneous tissues due to the nonlinear nature of scatter~\cite{ref23}. Deep
learning-based approaches attempt to learn the mappings from artifact-laden to artifact-free CBCT
images using paired datasets. These models operate in either the projection or image
domain~\cite{ref15,ref16}, and have demonstrated effectiveness in correcting
scatter~\cite{ref16}, streaking~\cite{ref15}, and metal artifacts~\cite{ref24}. However, they often
struggle to comprehensively address multiple artifact sources, which remains a shared limitation
across all artifact correction methods.

The sCT generation approaches leverage image prior information from the pCT to map CBCT voxels
to the target pCT via deep learning models including U-net and
GANs~\cite{ref6,ref17,ref19,ref20,ref21}. Denoising diffusion models have recently emerged as a
powerful generative framework that produced state-of-the-art image quality in medical imaging
including the CBCT-based sCT generation~\cite{ref25,ref26,ref27}. However, the denoising
diffusion models typically rely on a long sequence of iterative updates to progressively transform
an initial random noise sample into a realistic image---a procedure often referred to as the
sampling chain. At each iteration, the generated image is gradually refined toward higher fidelity.
While this iterative mechanism underpins the strong generative performance of diffusion models,
it also imposes a substantial computational burden: hundreds or even thousands of steps are often
required to obtain a single high-quality sample. Consequently, the sampling process becomes both
time-consuming and resource-intensive, which may constitute a practical bottleneck for applications
that demand rapid image synthesis~\cite{ref28}.

In this study, we propose a conditional flow matching framework for efficient sCT generation from
CBCT images. The model is trained using paired CBCT and deformed planning MDCT (dpCT
henceforth) images, where the dpCT serves as the image label and CBCT acts as the conditioning
image input. During inference, a white Gaussian noise sample is transformed in several steps into
an sCT image conditioned on the CBCT. With only 5--20 sampling steps, our method achieves
comparable performance to a conventional DDPM with 1000 steps, while substantially reducing
inference time and computational cost. Experimental results on brain, head-and-neck (HN), and
lung datasets demonstrate that the proposed method can efficiently generate sCT with improved HU
accuracy and reduced artifacts, thereby supporting the feasibility of direct CBCT-based ART.

% ================================================================
\section{Materials and Methods}
% ================================================================

\subsection{Image Acquisition and Preprocessing}

In this study, the CBCT and pCT images were acquired from 41 brain, 47 HN patients, and 37 lung
patients at our institutional center. For the brain cohort, training and testing datasets consist
of 4682 slices from 30 patients and 500 slices randomly selected from 11 patients, respectively.
In the HN patient study, 4314 slices from 37 patients were used for training, with 500 random
slices from other 10 patients were used for testing. For the lung cohort, the training set included
3216 slices from 27 patients and the testing set consist of 500 random slices from other 10
patients. All CT images were acquired at 120 kVp using the Siemens SOMATOM Definition AS
scanner. All CBCT images were acquired using the Varian TrueBeam system. For the brain and HN
cohort, CBCT images were acquired at 100 kVp, while the lung CBCT images were acquired at 125
kVp. Both CT and CBCT volumes had isotropic voxel dimensions of
$1.0\!\times\!1.0\!\times\!1.0$~mm$^{3}$. The pCT images were deformably registered and
resampled to align with the original CBCT images using a commercial software (Velocity AI version
3.2.1), ensuring matched volume size and voxel spacing across image pairs. The body contours
extracted from both modalities were combined and applied uniformly to the corresponding CBCT and
pCT scans. Prior to inputting into the neural networks, images were cropped to
$256\!\times\!256$ pixels (brain and HN patients) and $512\!\times\!272$ (lung patients) and
intensity normalized to the range of $[-1,\,1]$.

\subsection{Flow Matching Model}

Let $\mathbb{R}^{d}$ denote the data space, where each data point is given by
$x = (x_1, \ldots, x_d) \in \mathbb{R}^{d}$. The probability density path
$p : [0,1] \to \mathbb{R}^{d} \to \mathbb{R}_{>0}$ is a family of time-dependent probability
density functions that satisfies $\int p_t(x)\,dx = 1$. A time-dependent vector field
$v : [0,1] \to \mathbb{R}^{d} \to \mathbb{R}^{d}$ can be used to construct a time-dependent
diffeomorphic map, known as a flow,
$\psi : [0,1] \to \mathbb{R}^{d} \to \mathbb{R}^{d}$, defined by the ordinary differential
equation (ODE):
\begin{align}
    \frac{d}{dt}\psi_t(x) &= v_t(\psi_t(x)) \label{eq:ode1} \\
    \psi_0(x) &= x \label{eq:ode2}
\end{align}

Prior work proposed parameterizing the vector field $v_t$ with a neural network
$v_{t,\theta}(x)$, where $\theta \in \mathbb{R}^{p}$ are trainable parameters. This yields a
deep parametric model of the flow $\phi_t$~\cite{ref29}, referred to as a continuous normalizing
flow (CNF). The CNF framework allows a simple density $p_0$ (e.g., Gaussian noise) to be
progressively transformed into a complex target distribution $p_1$ via the push-forward relation
\begin{equation}
    p_t = [\psi_t]_* p_0, \label{eq:pushforward}
\end{equation}
where the operator $*$ is explicitly given by
\begin{equation}
    [\psi_t]_* p_0(x) = p_0\!\left(\psi_t^{-1}(x)\right)
    \det\!\left[\frac{\partial \psi_t^{-1}}{\partial x}(x)\right]. \label{eq:pushforward_explicit}
\end{equation}

However, the practical use of CNFs as generative models has been significantly limited by the high
computational cost associated with maximum likelihood training. To overcome this drawback, the
flow matching has recently been proposed as a viable alternative. The key observation underlying
this approach is that any time-dependent vector field
$v_t : [0,1] \to \mathbb{R}^{d} \to \mathbb{R}^{d}$ can be associated with the temporal
evolution of a prescribed density path
$p_t : [0,1] \to \mathbb{R}^{d} \to [0,\infty)$ through the fundamental continuity
equation~\cite{ref30}
\begin{equation}
    \frac{\partial p_t(x)}{\partial t} = -\nabla \cdot \bigl(p_t(x)\,v_t(x)\bigr), \label{eq:continuity}
\end{equation}
where $\nabla\cdot$ denotes the divergence operator. This equation ensures that the probability
mass is conserved throughout the transformation process.

The velocity field $v_t(x)$ is learned by a neural network $v_{t,\theta}(x)$ that minimizes the
following objective~\cite{ref31}:
\begin{equation}
    \mathcal{L}_{\mathrm{FM}} =
    \mathbb{E}_{t \sim \mathcal{U}[0,1],\; x_t \sim p_t}
    \Bigl[\bigl\|v_{t,\theta}(x_t) - v_t(x_t)\bigr\|^2\Bigr]. \label{eq:FM}
\end{equation}

In essence, the flow matching (FM) loss enforces the regression of the target vector field $v_t$
by a neural network parameterization $v_{t,\theta}$. When the loss converges to zero, the learned
CNF model recovers the desired density path $p_t(x)$. However, the ground truth of marginal
vector field $v_t$ is unknown in general, making the optimization problem above intractable.
Instead, prior works showed that $v_t$ could be constructed via superimposing conditional vector
fields $v_t(\cdot|x_1)$ conditioned on available training data $x_1 \sim p_1$. Then the
conditional flow matching (CFM) objective is given as~\cite{ref31}
\begin{equation}
    \mathcal{L}_{\mathrm{CFM}} =
    \mathbb{E}_{t \sim \mathcal{U}[0,1],\; x_1 \sim p_1,\; x_t \sim p_t(\cdot|x_1)}
    \Bigl[\bigl\|v_{t,\theta}(x_t) - v_t(x_t|x_1)\bigr\|^2\Bigr]. \label{eq:CFM}
\end{equation}

In expectation, the minimization of the CFM objective is equivalent to minimizing the FM
objective. As a result, we can train a CNF to generate the marginal probability path $p_t$,
which approximates the unknown data distribution $p_1$ at $t=1$ without requiring explicit
knowledge of either the marginal path $p_t$ or the corresponding marginal vector field.

The CFM objective works for any conditional probability path and conditional vector field. One of
the choices is known as the conditional optimal-transport (OT) path~\cite{ref32}, which can be
explicitly expressed as
\begin{equation}
    x_t = t \cdot x_1 + (1-t) \cdot x_0. \label{eq:OTpath}
\end{equation}
Then we get the implementation of FM as
\begin{equation}
    \mathcal{L}_{\mathrm{CFM}}^{\mathrm{OT}} =
    \mathbb{E}_{t \sim \mathcal{U}[0,1],\; x_0 \sim p_0,\; x_1 \sim p_1}
    \Bigl[\bigl\|v_{t,\theta}(x_t) - (x_1 - x_0)\bigr\|^2\Bigr]. \label{eq:OT_CFM}
\end{equation}

\begin{figure}[htbp]
    \centering
    % Replace with actual figure file, e.g.:
    \includegraphics[width=\textwidth]{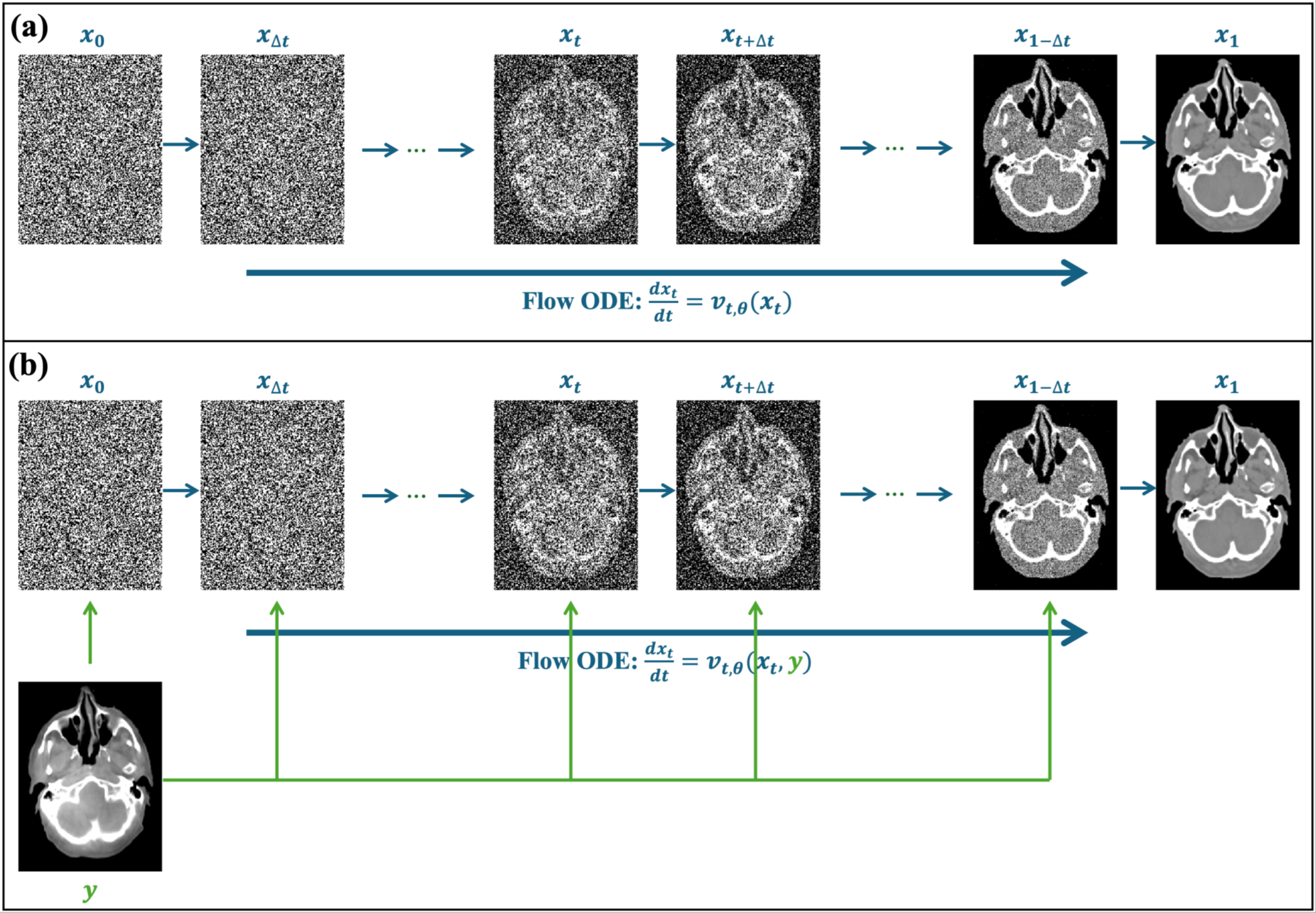}
    \caption{Workflows of (a) the standard flow matching model (Gaussian w/o Concat) and (b) the
    proposed conditional flow matching model (Gaussian w/ Concat) for synthetic CT generation
    from CBCT.}
    \label{fig:workflow}
\end{figure}

The workflow of the flow-matching model is shown in Figure~\ref{fig:workflow}.

\subsection{Sampling Process}

After training the flow-based model, sample generation is performed by integrating the flow ODE
forward in time. Recall that the dynamics of the learned flow are governed by
\begin{align}
    \frac{d}{dt}\psi_t(x) &= v_t(\psi_t(x)), \label{eq:samp_ode1}\\
    \psi_0(x) &= x, \label{eq:samp_ode2}
\end{align}
where the initial condition $x \sim p_0$ is drawn from a simple prior distribution such as a
standard Gaussian. The final state $\phi_1(x)$ provides a sample from the learned approximation
of the target data distribution $p_1$.

To approximate the continuous dynamics, we discretize the unit interval $[0,1]$ into $N$ equal
steps. Denoting the step size by $\Delta t = 1/N$ and the discretization grid by
$t_k = k\,\Delta t$ for $k = 0, \ldots, N$, we introduce the sequence $\{x_k\}_{k=0}^{N}$ to
represent the numerical trajectory, where $x_k \approx \psi_{t_k}(x)$.

Among the various numerical integrators available, we adopt the explicit Euler method for its
simplicity and efficiency. Concretely, the iteration proceeds as
\begin{equation}
    x_{k+1} = x_k + \Delta t\; v_{t_k}(x_k), \qquad x_0 \sim p_0. \label{eq:euler}
\end{equation}

This recursive scheme propagates the initial noise sample $x_0$ step by step until $x_N$, which
serves as an approximation to $\psi_1(x)$.

The Euler update rule may be interpreted as a sequence of incremental transport
operations~\cite{ref33}, where each step pushes the current point $x_k$ along the learned
velocity field $v_{t_k}$. Intuitively, this process gradually deforms the simple reference
distribution $p_0$ into the more complex data distribution $p_1$. From this perspective, the
sampling stage realizes a discrete-time approximation of the continuous push-forward relation
\begin{equation}
    p_t = [\psi_t]_* p_0, \label{eq:pushforward2}
\end{equation}
where the discretization is implemented via the iterative map.

In practice, the choice of the number of steps $N$ directly affects the trade-off between
computational efficiency and fidelity of the generated samples. A larger $N$ leads to a finer
discretization of the dynamics and a more faithful approximation of the target distribution, at
the expense of increased computational cost. On the other hand, a smaller $N$ yields faster
sampling but may introduce noticeable discretization bias. We observe empirically that the learned
flows remain stable under moderate step sizes, and high-quality samples can often be obtained
without requiring an excessively large $N$. This makes the Euler solver a practical and widely
applicable choice for sampling in flow matching models~\cite{ref31,ref32}.

\subsection{Flow Matching Model for Synthetic CT Generation}

The original flow matching model is designed as an unsupervised framework for unconditional image
generation and is therefore not directly applicable to tasks requiring semantic control over the
output~\cite{ref34}. In contrast, CT synthesis from CBCT falls under the category of conditional
image-to-image generation, where the generated sCT must correspond to a specific input CBCT
rather than being a random sample from the target CT distribution. To address this, various
conditioning mechanisms have been proposed to guide flow matching models for controlled
generation~\cite{ref35,ref36,ref37}.

In this study, we adopt a widely used strategy in conditional generative modeling, where the CBCT
image input ($y$) is concatenated with the flow sample ($x_t$) along the channel as a static
guidance during the velocity field approximation~\cite{ref25}. With the introduced conditioning
input, the velocity estimator in flow matching changes from $v_{t,\theta}(x_t)$ to
$v_{t,\theta}(x_t, y)$. The modified workflow was shown in Figure~\ref{fig:workflow}(b). The
training and sampling algorithms of the conditional flow matching model for CBCT-based synthetic
CT are summarized in Algorithms~\ref{alg:training} and~\ref{alg:sampling}.

\begin{algorithm}[htbp]
\caption{Flow Matching Training Procedure}
\label{alg:training}
\begin{algorithmic}[1]
\While{not converged}
    \State Sample $(y, x_1)$ from paired (CBCT, dpCT) dataset
    \State Sample $t$ from $\mathcal{U}[0,1]$
    \State Sample $x_0$ from Gaussian distribution
    \State Set $x_t = t \cdot x_1 + (1-t) \cdot x_0$
    \State Compute loss $\mathcal{L}_{\mathrm{CFM}}^{\mathrm{OT}} =
           \bigl\|v_{t,\theta}(x_t, y) - (x_1 - x_0)\bigr\|^2$
    \State Update $\theta$ via gradient descent on $\mathcal{L}_{\mathrm{CFM}}^{\mathrm{OT}}$
\EndWhile
\end{algorithmic}
\end{algorithm}

\begin{algorithm}[htbp]
\caption{Euler Sampling}
\label{alg:sampling}
\begin{algorithmic}[1]
\State Set $t = 0$
\State Set step size $\Delta t = \frac{1}{N}$
\State Sample $x_0$ from Gaussian distribution
\For{$k = 0$ to $N-1$}
    \State $x_{t+\Delta t} = x_t + \Delta t \cdot v_{t,\theta}(x_t, y)$
    \State $t = t + \Delta t$
\EndFor
\State \Return $x_1$
\end{algorithmic}
\end{algorithm}

\subsection{Comparison Studies}

In the comparison studies, we include as a baseline the results obtained from a conditional
diffusion probabilistic model (cDDPM) trained with 1000 denoising steps~\cite{ref25}. By
comparing our flow matching formulations against the 1000-step cDDPM baseline, we are able to
assess the behavior of flow-based methods in relation to a well-established diffusion approach,
thereby situating our study within the broader context of generative modeling for CBCT-to-CT
translation.

A major difference between the flow matching framework and the denoising diffusion models is the
remarkable flexibility it offers in specification of the initial distribution. Traditional
generative models, such as the variational autoencoders or denoising diffusion-based approaches,
typically rely on simple priors (e.g., isotropic Gaussian) as the starting point of the generative
process in order to facilitate mathematical tractability. The flow matching, by contrast, imposes
no such restriction: the initial distribution $p_0$ can, in principle, be chosen freely, as long
as it is amenable to sampling. This property provides substantial modeling flexibility, since it
allows the generative process to begin from a distribution that is already semantically or
structurally close to the target distribution.

In this work, we take advantage of this property for methodological comparison by directly
initializing the probability path from the empirical distribution of CBCT images, rather than
adopting a synthetic prior such as Gaussian noise. Starting from the CBCT samples allows us to
investigate how flow matching behaves when the source distribution is already structurally similar
to the target CT distribution, rather than being completely unrelated. We emphasize that this
choice of design is not necessarily intended to yield superior performance; rather, it highlights a
distinctive modeling option enabled by flow matching. By examining both Gaussian-to-CT and
CBCT-to-CT formulations, we aim to provide a clearer picture of how the choice of initial
distribution may influence the training dynamics and the generative mapping learned by the model.

\begin{figure}[htbp]
    \centering
    \includegraphics[width=\textwidth]{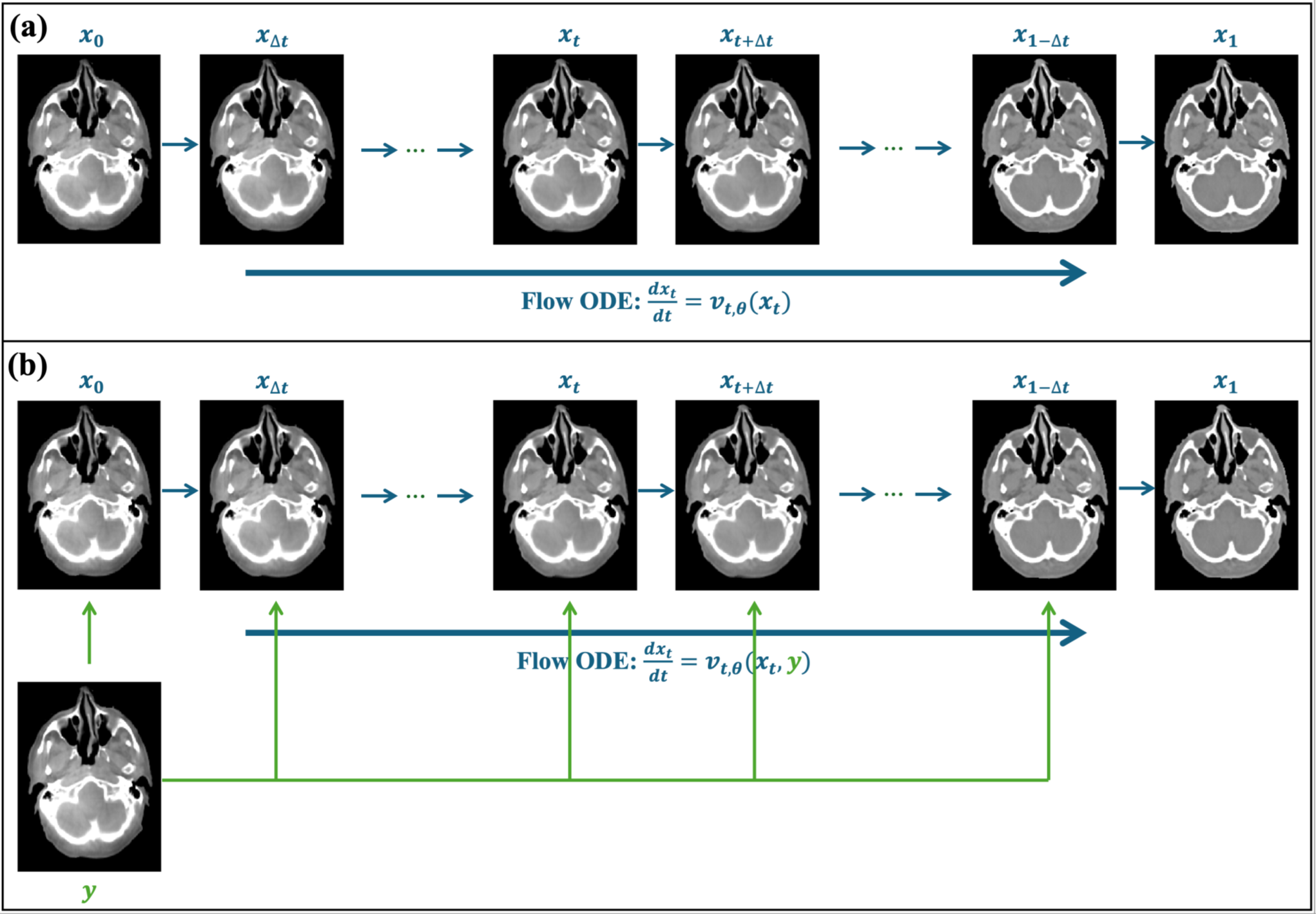}
    \caption{Workflows of (a) CBCT w/o Concat and (b) CBCT w/ Concat flow matching models for
    synthetic CT generation from CBCT.}
    \label{fig:comparison}
\end{figure}

To further examine the role of the initial distribution and conditional information in flow
matching, we adopt two complementary strategies for
comparison~\cite{ref38,ref39}. In the first strategy (\textbf{CBCT w/o Concat}), we directly
construct the flow from the CBCT distribution to the CT distribution, thereby treating CBCT as
the starting point of the probability path, as shown in Figure~\ref{fig:comparison}(a). In the
second strategy (\textbf{CBCT w/ Concat}), we build upon this formulation by additionally
concatenating the CBCT image at each time step as an explicit conditioning input to the neural
vector field, as depicted in Figure~\ref{fig:comparison}(b). This conditional variant allows the
model to incorporate structural information from CBCT more explicitly during the flow evolution,
offering a different perspective on how the source modality can be utilized in the generative
process. Together, these two strategies provide a basis for systematically investigating how the
choice of initialization and conditioning influences the learned mapping from CBCT to CT.
The training and inference stages of these two methods are summarized in Algorithms~\ref{alg:cbct_nowoc_train}--\ref{alg:cbct_wc_samp}.

\begin{algorithm}[htbp]
\caption{Flow Matching Training Procedure of (CBCT w/o Concat) Strategy}
\label{alg:cbct_nowoc_train}
\begin{algorithmic}[1]
\While{not converged}
    \State Sample $(x_0, x_1)$ from paired (CBCT, dpCT) dataset
    \State Sample $t$ from $\mathcal{U}[0,1]$
    \State Set $x_t = t \cdot x_1 + (1-t) \cdot x_0$
    \State Compute loss $\mathcal{L}_{\mathrm{CFM}}^{\mathrm{OT}} =
           \bigl\|v_{t,\theta}(x_t) - (x_1 - x_0)\bigr\|^2$
    \State Update $\theta$ via gradient descent on $\mathcal{L}_{\mathrm{CFM}}^{\mathrm{OT}}$
\EndWhile
\end{algorithmic}
\end{algorithm}

\begin{algorithm}[htbp]
\caption{Euler Sampling of (CBCT w/o Concat) Strategy}
\label{alg:cbct_nowoc_samp}
\begin{algorithmic}[1]
\State Set $t = 0$
\State Set step size $\Delta t = \frac{1}{N}$
\State Sample $x_0$ from CBCT distribution
\For{$k = 0$ to $N-1$}
    \State $x_{t+\Delta t} = x_t + \Delta t \cdot v_{t,\theta}(x_t)$
    \State $t = t + \Delta t$
\EndFor
\State \Return $x_1$
\end{algorithmic}
\end{algorithm}

\begin{algorithm}[htbp]
\caption{Flow Matching Training Procedure of (CBCT w/ Concat) Strategy}
\label{alg:cbct_wc_train}
\begin{algorithmic}[1]
\While{not converged}
    \State Sample $(x_0, x_1)$ from paired (CBCT, dpCT) dataset
    \State Sample $t$ from $\mathcal{U}[0,1]$
    \State Set $x_t = t \cdot x_1 + (1-t) \cdot x_0$
    \State Compute loss $\mathcal{L}_{\mathrm{CFM}}^{\mathrm{OT}} =
           \bigl\|v_{t,\theta}(x_t, x_0) - (x_1 - x_0)\bigr\|^2$
    \State Update $\theta$ via gradient descent on $\mathcal{L}_{\mathrm{CFM}}^{\mathrm{OT}}$
\EndWhile
\end{algorithmic}
\end{algorithm}

\begin{algorithm}[htbp]
\caption{Euler Sampling of (CBCT w/ Concat) Strategy}
\label{alg:cbct_wc_samp}
\begin{algorithmic}[1]
\State Set $t = 0$
\State Set step size $\Delta t = \frac{1}{N}$
\State Sample $x_0$ from CBCT distribution
\For{$k = 0$ to $N-1$}
    \State $x_{t+\Delta t} = x_t + \Delta t \cdot v_{t,\theta}(x_t, x_0)$
    \State $t = t + \Delta t$
\EndFor
\State \Return $x_1$
\end{algorithmic}
\end{algorithm}

To systematically evaluate these strategies, we examine the quality of generated CT images under
different discretization granularities in the sampling stage. Specifically, for each strategy we
report results obtained using Euler integration with $N \in \{5,10,20\}$ steps. This allows us to
assess how the number of integration steps influences the fidelity and stability of the synthesized
images, and to compare the relative robustness of the two flow matching formulations under coarse
versus fine discretization.

\subsection{Implementation and Evaluation}

A U-Net architecture with residual blocks, attention modules, and time embeddings was adopted for
vector field estimation at each step. All experiments in this work were conducted using PyTorch
1.12 framework on an NVIDIA RTX Ada 6000 GPU with 48 GB memory. Training was terminated after
$1 \times 10^6$ iterations, which took approximately 10 hours for the brain and HN studies and
30 hours for the lung study. The sampling stage of a single slice consumed 29/86 seconds (brain
and HN patients/lung patients) using the baseline cDDPM method with 1000 time steps, while it
only took 0.09/0.25 seconds, 0.21/0.64 seconds, and 0.49/1.41 seconds using the flow matching
model with 5, 10, and 20 time steps. The sampling time was averaged over all 500 testing slices.

To evaluate the performance of the generated sCT images, we computed the mean absolute error
(MAE), peak signal-to-noise ratio (PSNR), and normalized cross-correlation (NCC) with respect to
the reference images (dpCT). These metrics are mathematically defined as follows:
\begin{align}
    \mathrm{MAE} &= \frac{1}{N}\sum_{m,n} \bigl|\mathrm{sCT}(m,n) - \mathrm{REF}(m,n)\bigr|
    \label{eq:MAE}\\[6pt]
    \mathrm{PSNR} &= 10 \cdot \log_{10}\!\left(
        \frac{\mathrm{MAX}^2}{\dfrac{1}{N}\sum_{m,n}\bigl(\mathrm{sCT}(m,n)-\mathrm{REF}(m,n)\bigr)^2}
    \right) \label{eq:PSNR}\\[6pt]
    \mathrm{NCC} &= \frac{1}{N}\sum_{m,n}
    \frac{\bigl(\mathrm{sCT}(m,n)-\overline{\mathrm{sCT}}\bigr)
          \bigl(\mathrm{REF}(m,n)-\overline{\mathrm{REF}}\bigr)}
         {\sigma_{\mathrm{sCT}}\,\sigma_{\mathrm{REF}}}
    \label{eq:NCC}
\end{align}
where $\mathrm{sCT}(m,n)$ and $\mathrm{REF}(m,n)$ denote the pixel values at location $(m,n)$
in the sCT and reference images, and $N$ is the total number of pixels of the image. $\mathrm{MAX}$
is the maximum pixel intensity of the image. $\overline{\mathrm{sCT}}$ and $\overline{\mathrm{REF}}$
are the mean intensities, and $\sigma_{\mathrm{sCT}}$ and $\sigma_{\mathrm{REF}}$ are the standard
deviations of the sCT and reference images, respectively.

The MAE quantifies the voxel-wise HU discrepancy between the sCT and CT. PSNR evaluates the
global intensity consistency, and NCC measures the structural similarity between the two images.

% ================================================================
\section{Results}
% ================================================================

\subsection{Improvement of Visual Quality}

\begin{figure}[htbp]
    \centering
    \includegraphics[width=\textwidth]{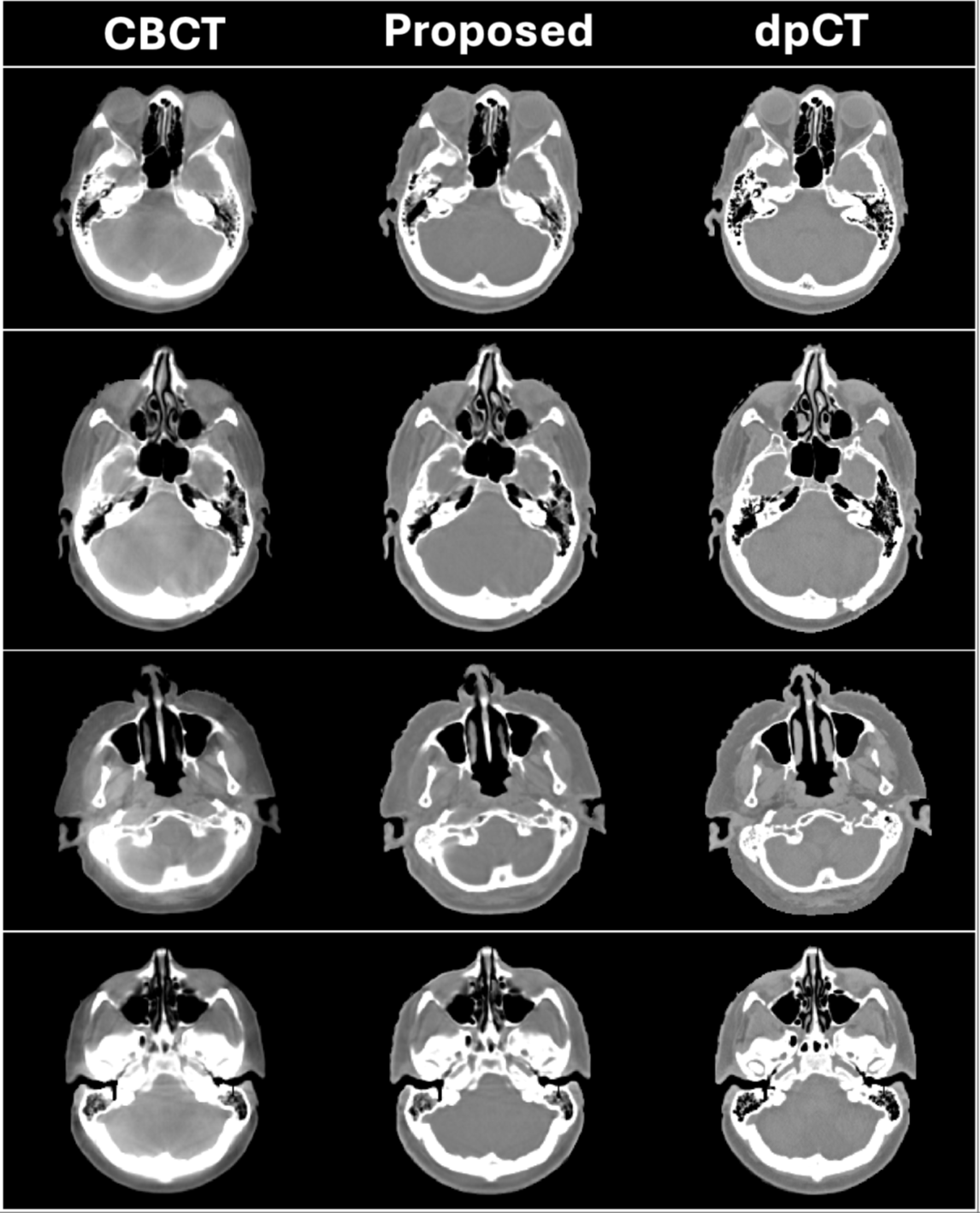}
    \caption{Visual effects of artifacts reduction in the study of brain patient. Left to right:
    CBCT, sCT generated by the proposed flow-matching model (5 step), and reference dpCT,
    respectively. Display window: [-500~500] HU.}
    \label{fig:brain}
\end{figure}

The visual effect of artifact correction using the proposed method is illustrated in
Figure~\ref{fig:brain} over the brain patient cases. Pronounced beam-hardening artifacts,
originating from the high-density bones, are evident in all the CBCT images, manifesting as the
inhomogeneity in intensity within the brain region and blurred boundaries between the bone and
adjacent soft tissue. In contrast, the CT images generated by our method exhibit a substantial
reduction of those artifacts while faithfully preserving fine anatomical structures, including the
contrast agent and the maxillary sinus. This improvement is expected to enhance the accuracy of
downstream tasks such as organ segmentation.

\begin{figure}[htbp]
    \centering
    \includegraphics[width=\textwidth]{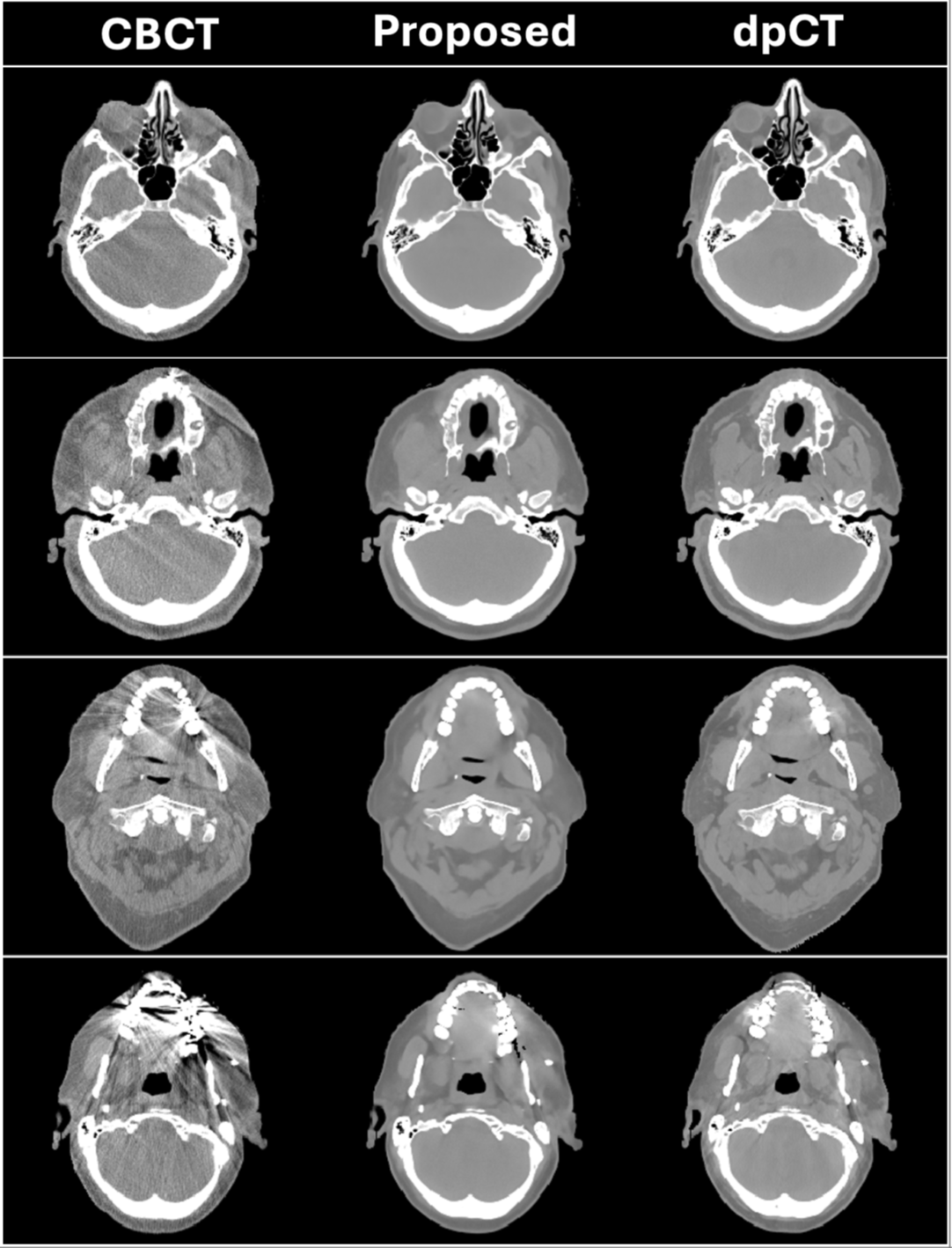}
    \caption{Visual effects of artifacts correction in the study of HN patient. Left to right:
    CBCT, sCT generated by the proposed flow-matching model (5 step), and reference dpCT,
    respectively. Display window: [-500~500] HU.}
    \label{fig:hn}
\end{figure}

\begin{figure}[htbp]
    \centering
    \includegraphics[width=\textwidth]{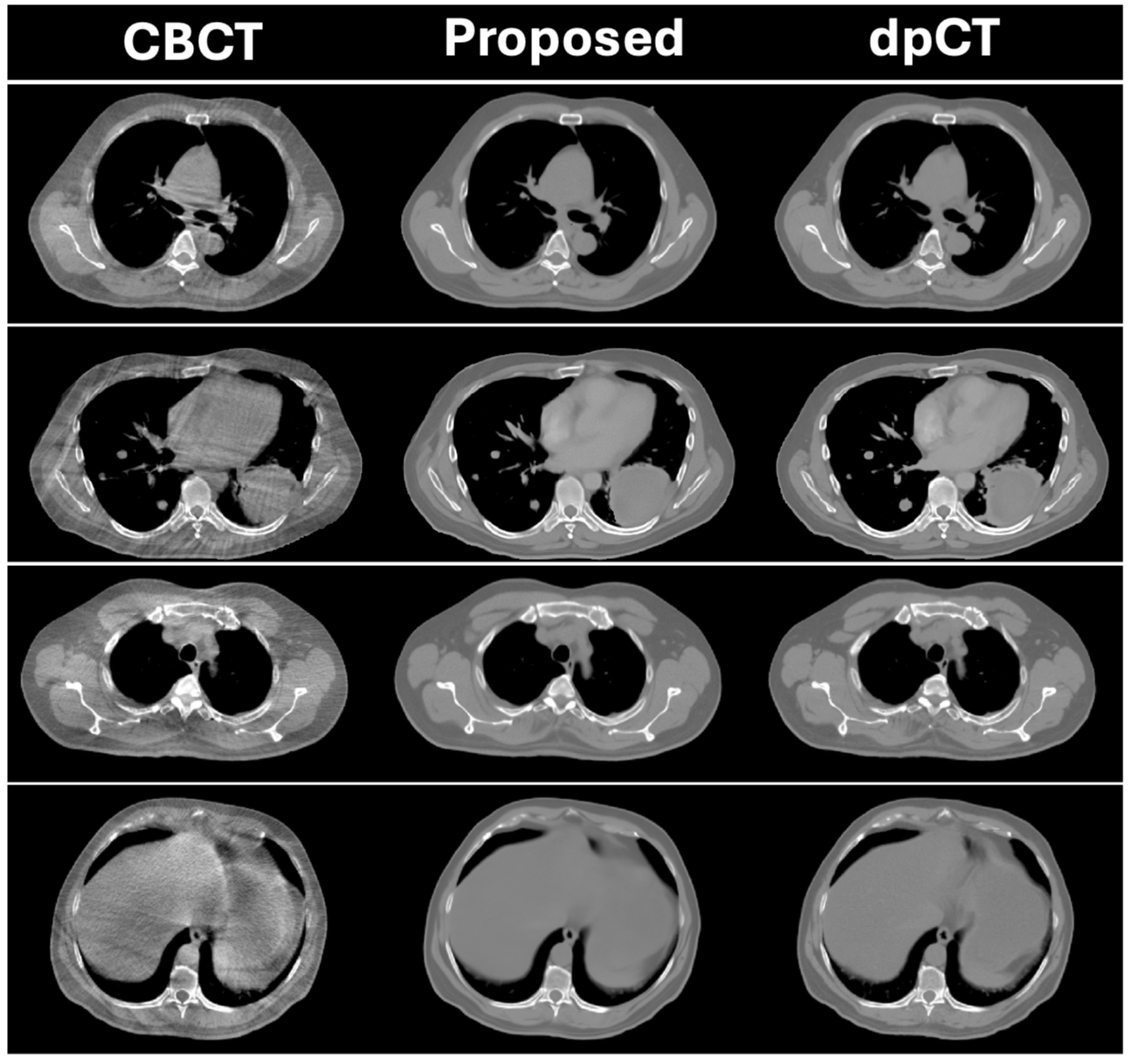}
    \caption{Visual effects of artifacts correction in the lung cohort. Left to right: CBCT, sCT
    generated by the proposed flow-matching model (5 step), and reference dpCT, respectively.
    Display window: [-500~500] HU.}
    \label{fig:lung}
\end{figure}

Figures~\ref{fig:hn} and~\ref{fig:lung} illustrate the visual improvements achieved by the
proposed sCT method in the studies of HN patient and lung patient. The CBCT images were heavily
degraded by streaking artifacts caused by motion and the presence of metallic implants. For the
first two slices, the proposed approach produced artifact-free reconstructions that closely
resemble the corresponding dpCT images. Furthermore, for severely degraded slices affected by
strong metal artifacts, such as the last two slices, the sCT images generated by the proposed
method exhibit effective artifact suppression that are compared to the dpCT images.

\subsection{Quantitative Analysis}

Table~\ref{tab:quantitative} tabulates the quantitative results for acquired CBCT and synthetic
CT using the proposed method. For the brain patient study, the flow-matching model improved the
HU accuracy from MAE of 40.63 HU for CBCT to 26.02 HU, 25.84 HU, and 26.31 HU with 5, 10, and
20 time steps. PSNR was improved from 27.87 dB for CBCT and 32.35 dB, 32.43 dB, and 32.15 dB
for sCT with 5, 10, and 20 time steps. The NCC was 0.98 for CBCT and 0.99 for all the sCT
results. For the HN patient study, MAE decreased from 38.99 HU to 33.17 HU, 33.02 HU, and 33.42
HU for sCT using 5, 10, and 20 steps. PSNR was slightly improved while NCC kept the same. For
the lung patient study, MAE decreased from 32.90 HU to 25.09 HU, 24.20 HU, and 24.34 HU for sCT
using 5, 10, and 20 steps. PSNR and NCC were slightly improved.

\begin{sidewaystable}[htbp]
\centering
\caption{Quantitative metrics (mean $\pm$ std) of the sCT using different methods.}
\label{tab:quantitative}
\small
\setlength{\tabcolsep}{5pt}
\begin{tabular}{p{3.0cm}ccccccccc}
\toprule
\multirow{2}{*}{Method} &
\multicolumn{3}{c}{MAE (HU)} &
\multicolumn{3}{c}{PSNR (dB)} &
\multicolumn{3}{c}{NCC} \\
\cmidrule(lr){2-4}\cmidrule(lr){5-7}\cmidrule(lr){8-10}
& Brain & HN & Lung & Brain & HN & Lung & Brain & HN & Lung \\
\midrule
CBCT
  & $40.63\pm12.71$ & $38.99\pm14.07$ & $32.90\pm11.45$
  & $27.87\pm2.20$  & $27.00\pm1.98$  & $30.48\pm2.36$
  & $0.98\pm0.01$   & $0.98\pm0.01$   & $0.98\pm0.01$ \\
cDDPM (1000-Step)
  & $25.99\pm11.84$ & $32.56\pm12.86$ & $24.68\pm8.27$
  & $30.49\pm3.73$  & $27.65\pm2.41$  & $31.78\pm2.14$
  & $0.99\pm0.01$   & $0.98\pm0.01$   & $0.99\pm0.01$ \\
Proposed (5-Step)
  & $26.02\pm8.65$  & $33.17\pm9.74$  & $25.09\pm6.60$
  & $32.35\pm3.42$  & $28.68\pm2.32$  & $32.81\pm2.35$
  & $0.99\pm0.01$   & $0.98\pm0.01$   & $0.99\pm0.01$ \\
Proposed (10-Step)
  & $25.84\pm8.60$  & $33.02\pm9.73$  & $25.20\pm6.58$
  & $32.43\pm3.41$  & $28.78\pm2.35$  & $32.85\pm2.43$
  & $0.99\pm0.01$   & $0.98\pm0.01$   & $0.99\pm0.01$ \\
Proposed (20-Step)
  & $26.31\pm8.63$  & $33.42\pm9.77$  & $24.34\pm6.55$
  & $32.15\pm3.33$  & $28.61\pm2.31$  & $32.79\pm2.31$
  & $28.61\pm2.31$  & $0.99\pm0.01$   & $0.98\pm0.01$ \\
CBCT w/o Concat (5-Step)
  & $27.30\pm8.80$  & $35.54\pm10.43$ & $25.01\pm8.29$
  & $31.18\pm2.99$  & $28.26\pm2.22$  & $32.93\pm2.62$
  & $0.99\pm0.01$   & $0.98\pm0.01$   & $0.99\pm0.01$ \\
CBCT w/o Concat (10-Step)
  & $27.61\pm8.84$  & $35.99\pm10.67$ & $25.27\pm8.43$
  & $31.07\pm2.97$  & $28.20\pm2.21$  & $32.89\pm2.62$
  & $0.99\pm0.01$   & $0.98\pm0.01$   & $0.99\pm0.01$ \\
CBCT w/o Concat (20-Step)
  & $27.79\pm8.86$  & $36.18\pm10.68$ & $25.47\pm8.54$
  & $31.01\pm2.96$  & $28.15\pm2.19$  & $32.86\pm2.63$
  & $0.99\pm0.01$   & $0.98\pm0.01$   & $0.99\pm0.01$ \\
CBCT w/ Concat (5-Step)
  & $27.18\pm10.02$ & $35.26\pm12.69$ & $25.68\pm9.34$
  & $30.57\pm2.67$  & $28.02\pm2.23$  & $32.02\pm2.65$
  & $0.99\pm0.01$   & $0.98\pm0.01$   & $0.99\pm0.01$ \\
CBCT w/ Concat (10-Step)
  & $27.11\pm10.04$ & $35.60\pm12.72$ & $25.31\pm9.67$
  & $30.56\pm2.68$  & $27.96\pm2.22$  & $32.13\pm2.63$
  & $0.99\pm0.01$   & $0.98\pm0.01$   & $0.99\pm0.01$ \\
CBCT w/ Concat (20-Step)
  & $28.21\pm9.16$  & $36.04\pm12.89$ & $25.57\pm9.42$
  & $30.12\pm2.63$  & $27.87\pm2.22$  & $32.07\pm2.61$
  & $0.99\pm0.01$   & $0.98\pm0.01$   & $0.99\pm0.01$ \\
\bottomrule
\end{tabular}
\end{sidewaystable}

\subsection{Comparison Studies}

\begin{figure}[htbp]
    \centering
    \includegraphics[width=\textwidth]{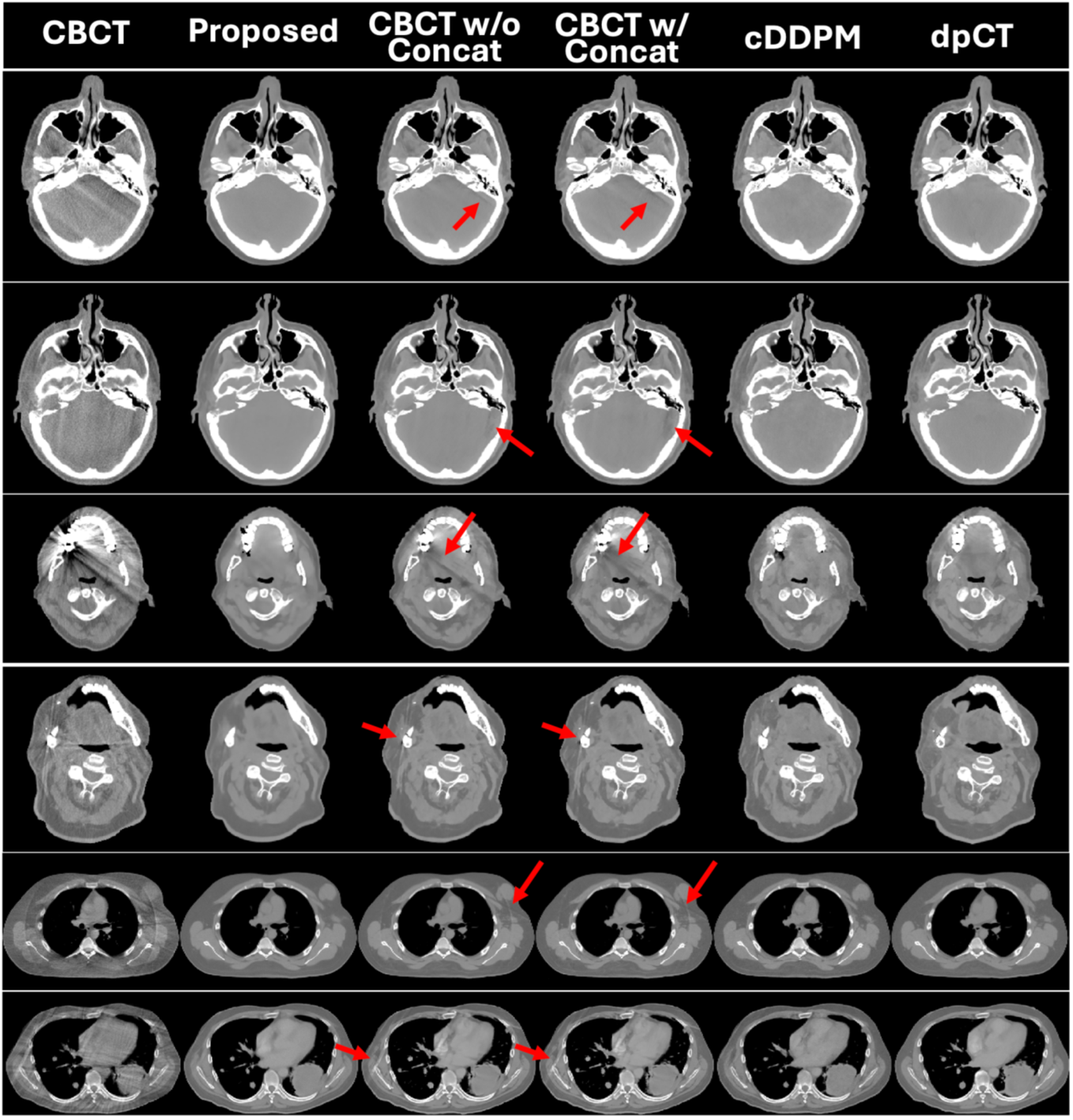}
    \caption{Results of sCT generated by different flow matching models (5 step) and baseline
    cDDPM (1000 step). Red arrows indicate the residual artifacts on sCT produced by compared
    flow matching models. Display window: [-500~500] HU.}
    \label{fig:comparison_results}
\end{figure}

The sCT images generated by different flow matching models (5 step) and baseline conditional DDPM
are summarized in Figure~\ref{fig:comparison_results}. Their quantitative evaluation metrics were
summarized in Table~\ref{tab:significance}. Compared to the sCT generated from the proposed
method, the results from comparing the flow matching models exhibited varying levels of residual
artifacts (indicated by the red arrows), particularly in the cases where the CBCT input contained
severe degradations such as metal-induced streaking and ring artifacts. A plausible explanation is
that, unlike the proposed method which starts from pure noise, the flow matching formulations are
initialized directly from heavily corrupted CBCT images, making it more challenging to suppress
artifacts that are already strongly embedded in the image domain. In addition, for each type of
flow matching model, the differences among results obtained with 5, 10, and 20 integration steps
were relatively minor, as can be observed from the quantitative metrics in Table~\ref{tab:significance}.

We conducted Welch's $t$-test to compare the results of the 5-step flow matching model with those
of the 1000-step cDDPM, 10-step flow matching, and 20-step flow matching models, as summarized in
Table~\ref{tab:significance}. The results indicate that the proposed method with 5, 10, and 20
steps exhibited no significant differences across all three metrics. In contrast, when compared
with the 1000-step cDDPM, the 5-step flow matching model showed no significant difference in MAE,
but achieved significantly better performance in both PSNR and NCC.

\begin{table}[htbp]
\centering
\caption{Results of significance tests ($p$-value).}
\label{tab:significance}
\small
\setlength{\tabcolsep}{5pt}
\begin{tabular}{p{3.2cm}ccccccccc}
\toprule
\multirow{2}{*}{Comparison} &
\multicolumn{3}{c}{MAE} &
\multicolumn{3}{c}{PSNR} &
\multicolumn{3}{c}{NCC} \\
\cmidrule(lr){2-4}\cmidrule(lr){5-7}\cmidrule(lr){8-10}
& Brain & HN & Lung & Brain & HN & Lung & Brain & HN & Lung \\
\midrule
Proposed (5-Step) vs.\ cDDPM (1000-Step)
  & 0.57 & 0.40 & 0.43 & $<$0.01 & $<$0.01 & $<$0.01 & $<$0.01 & $<$0.01 & $<$0.01 \\
Proposed (5-Step) vs.\ Proposed (10-Step)
  & 0.74 & 0.81 & 0.52 & 0.69 & 0.51 & 0.23 & 0.66 & 0.34 & 0.61 \\
Proposed (5-Step) vs.\ Proposed (20-Step)
  & 0.59 & 0.68 & 0.31 & 0.35 & 0.61 & 0.34 & 0.51 & 0.58 & 0.42 \\
\bottomrule
\end{tabular}
\end{table}

% ================================================================
\section{Discussion}
% ================================================================

This study formulated CBCT-based CT synthesis under the conditional flow matching framework. To
the best of our knowledge, this study represents the first attempt to perform CBCT-based CT
synthesis using a flow matching-based generative model. Unlike denoising diffusion model-based
methods, the proposed method efficiently generates enhanced-quality sCT images with significantly
reduced sampling steps, thus reducing computational cost while maintaining or even improving the
quality of generated sCT images.

One of the major limitations of this study is the supervised learning schemes employed for medical
image translation. The supervised learning method relies on exactly matched image pairs, which are
often unavailable in practice. For instance, it is infeasible to obtain paired CBCT--dpCT images
from the same patient both with and without metal artifacts. Moreover, in radiotherapy practice,
the anatomical changes such as tumor shrinkage, weight loss, or posture variation across treatment
sessions further increase the mismatch between pCT and CBCT, making precise registration
unattainable and thereby preventing the construction of exactly matched training pairs. In this
context, it should be noticed that the quantitative comparisons between sCT and dpCT should be
interpreted with caution, since the dpCT is not anatomically identical to the CBCT and may itself
contain artifacts, whereas the sCT may be closer to the ground truth image.

It is also of interest to further explore the feasibility of flow matching models for sCT
generation without CBCT and dpCT data pairs. There have been several recent attempts to leverage
the Bayesian formulations in conjunction with diffusion models to achieve unsupervised image
generation and reconstruction~\cite{ref26,ref40}. While these approaches provide elegant
approaches under the diffusion modeling framework, their underlying mathematical principles differ
fundamentally from those of flow matching. Moreover, the flow matching often operates with a
relatively small number of integration steps, which may further complicate the incorporation of
data consistency strategies. As a result, the Bayesian--diffusion strategies cannot be directly
transferred to the flow matching setting. Developing analogous formulations that are compatible
with the flow matching paradigm remains an open problem and will be an interesting direction for
future investigation.

In this work, the proposed flow-matching model is implemented on 2D slices and does not employ
any patch extraction strategy. While this choice of design simplifies training and reduces
computational cost, it inevitably limits the model's ability to fully capture 3D anatomical
context. It is therefore reasonable to expect that a 3D latent-space or patch-based
model~\cite{ref41}, which can exploit volumetric information and spatial correlations across
slices, would yield improved structural fidelity and more consistent preservation of anatomical
continuity. Incorporating such 3D modeling strategies represents a promising direction for future
work.

In this work, the sampling stage of the flow-based model was implemented using the explicit Euler
solver to numerically integrate the underlying ODE. The Euler integration is simple and efficient,
but it is also a first-order method and may introduce discretization bias while only a limited
number of steps are being employed. More advanced solvers, such as higher-order Runge--Kutta
methods, adaptive-step solvers, or predictor--corrector schemes, could potentially improve the
trade-off between computational cost and sample fidelity. A systematic investigation of how
different ODE solvers affect both the visual quality of the synthesized CT images and the
stability of the sampling process will be pursued in future work.

% ================================================================
\section{Conclusions}
% ================================================================

In this study, we proposed a conditional flow matching framework for synthetic CT directly from
CBCT inputs. The synthetic CT images demonstrated improved HU accuracy and exhibited
substantially fewer artifacts compared to the original CBCT images, which may facilitate more
direct CBCT-based organ segmentation and treatment replanning. Overall, the proposed method may
extend the quantitative utility of CBCT and strengthens its potential role in the clinical
workflow of direct CBCT-guided adaptive radiotherapy.

% ================================================================
% References
% ================================================================

\end{document}